\documentclass[twocolumn,floatfix,amsmath,amssymb]{revtex4-1}

\usepackage{graphicx}
\usepackage{dcolumn} 
\usepackage{bm}      
\usepackage[usenames,dvipsnames]{color}
\usepackage{ulem}    

\begin{document}
\title{Stably stratified turbulence in the presence of large-scale forcing}
\author{C. Rorai$^{1}$,  P.D. Mininni$^{2}$ and A. Pouquet$^{3,4}$}
\affiliation{
$^1$Nordita, Roslagstullsbacken 23, 106 91 Stockholm, Sweden; \\
$^2$Departamento de F\'\i sica, Facultad de Ciencias Exactas y Naturales, Universidad de Buenos Aires \\ 
        \& IFIBA, CONICET, Ciudad Universitaria, 1428 Buenos Aires, Argentina;\\
$^3$National Center for Atmospheric Research, P.O. Box 3000, Boulder CO 80307 USA; \\
$^4$Department of Applied Mathematics, University of Colorado, Boulder CO 80309-256 USA.
}

\begin{abstract}
We perform two high resolution direct numerical simulations of stratified turbulence for  Reynolds number equal to $Re\approx25000$ and Froude number respectively of $Fr\approx0.1$ and $Fr\approx0.03$. The flows are forced at large scale and discretized on an isotropic grid of $2048^3$ points. Stratification makes the flow anisotropic and introduces two extra characteristic scales with respect to homogeneous isotropic turbulence: the buoyancy scale, $L_B$, and the Ozmidov scale, $\ell_{oz}$. The former is related to the number of layers that the flow develops in the direction of gravity, the latter is regarded as the scale at which isotropy is recovered. The values of $L_B$ and $\ell_{oz}$ depend on the Froude number and their absolute and relative size affect the repartition of energy among Fourier modes in non easily predictable ways. By contrasting the behavior of the two simulated flows we identify some surprising similarities: after an initial transient the two flows evolve towards comparable values of the kinetic and potential enstrophy, and energy dissipation rate. This is the result of the Reynolds number being large enough in both flows for the Ozmidov scale to be resolved. When properly dimensionalized, the energy dissipation rate is compatible with atmospheric observations. Further similarities emerge at large scales: the same ratio between potential and total energy ($\approx 0.1$) is spontaneously selected by the flows, and slow modes grow monotonically in both regimes causing a slow increase of the total energy in time. The axisymmetric total energy spectrum shows a wide variety of spectral slopes as a function of the angle between the imposed stratification and the wave vector. One-dimensional energy spectra computed in the direction parallel to gravity are flat from the forcing up to buoyancy scale. At intermediate scales a $\sim k^{-3}$ parallel spectrum develops for the $Fr\approx 0.03$ run, whereas for weaker stratification, the saturation spectrum does not have enough scales to develop and instead one observes a power law compatible with Kolmogorov 
 scaling.
 Finally, the spectrum of helicity is flat until $L_B$, as observed in the nocturnal planetary boundary layer.
\end{abstract}

\pacs{47.55.Hd,  
	47.27.-i,  
	47.35.Bb,  
	47.27.ek } 
\maketitle

\section{Introduction} \label{sec:intro}
Geophysical fluid dynamics, as encountered in the atmosphere and the oceans, is at the center of our understanding and predicting capabilities in weather and climate. The modes that prevail in such systems are a mixture of nonlinear eddies and waves (for example, inertial waves when a solid-body rotation is considered, or internal gravity waves in a stratified flow). The nonlinear coupling between these modes leads to extreme events which are both sporadic and spatially localized, with steep gradients in the velocity and temperature or density fields, a phenomenon observed both in the stable planetary boundary layer \cite{lenschow_12}, as well as in high-resolution direct numerical simulations (DNS) of the Bousinesq equations \cite{rorai_14}. Nonlinear interactions are also associated with other phenomena, such as wave steepening and breaking, instabilities (as the Kelvin-Helmoltz instability, a familiar phenomenon in a turbulent stratified atmosphere), and turbulent cascades. As these interactions take place in a wide range of scales, and as timescales in geophysical flows are not homogeneous in scale, universality (as is the case in homogeneous isotropic turbulence) is not necessarily obtained in such complex flows. Numerous observations point to a variety of regimes, for example in the case of surface waves in the ocean, which steepen into well-observed rogue waves \cite{choi_05b, fedele_08, dysthe_08}. 

Whether the flow is dominated by (strong) waves or by eddies depends a priori on the relative values of the parameters that characterize a given flow. In the case of stratification, these are the period of gravity waves $\tau_\omega\approx 1/N$, and the turnover time of eddies $\tau_\textrm{NL}\approx L_F/u_\textrm{rms}$ (with $N$, $u_\textrm{rms}$, and $L_F$ respectively the 
Brunt-V\"ais\"al\"a frequency and the characteristic velocity and lengthscale of the flow). The fastest timescale is expected to dominate the dynamics. However, as already mentioned, these timescales are not homogeneous with the lengthscale, and even if the waves dominate over the eddies at large scales, the two timescales may become comparable at a smaller scale. For stratification, the scale at which these two characteristic times are equal is called the Ozmidov scale $\ell_\textrm{oz}=2 \pi/k_\textrm{oz}$, with $k_\textrm{oz} = (N^3/\epsilon)^{1/2}$ and with $\epsilon$ the energy dissipation rate. Beyond $\ell_\textrm{oz}$, isotropy and a classical Kolmogorov range is expected. The scale $\eta$ at which dissipation sets in (of the order of the meter or several centimeters in the atmosphere) marks the end of the turbulence regime. We thus propose in this paper to study forced stratified turbulence using DNS at a sufficiently high resolution of $2048^3$ points to be able to contrast the evolution of
 such flows in two cases, varying only the ratio $\tau_\omega/\tau_\textrm{NL}$ by a factor 3.

One can find many reviews concerning stratified turbulence (see, e.g., \cite{riley_rev_00, staquet_rev_02, mcwilliams_04, sagaut_cambon_08, waite_14}). Concentrating on more recent numerical work, a few concepts seem to emerge, among a variety of possible settings (two-dimensional or three-dimensional forcing, acting at large scale or at small scale, balanced or not, etc.) \cite{bartello_13, brethouwer_07, waite2011, kimura_12, almalkie_12, debruynkops_13, debruynkops_14}. Several large DNS also consider the stratified case in the presence of shear \cite{fritts_09a, chung_12} or of rotation \cite{smith_02, 4096_LS}.

In stratified turbulence, the kinetic energy undergoes a direct cascade to small scales, and its spectrum follows at sufficiently small scales a Kolmogorov-like law in terms of $k_{\perp}$ (i.e., of wavevectors perpendicular to gravity) \cite{lindborg_07}. At small scales  the buoyancy field is believed to follow an equivalent law, similar to that of a passive scalar, again in terms of $k_{\perp}$. However, the spectra in terms of $k_{\parallel}$ seem to follow a steeper $\sim k_\parallel^{-3}$ law, often called the saturation spectrum. Flat spectra at large scale, presumably larger than the buoyancy scale $L_B$, are also reported. Moreover, different simulations with varying configurations and parameters present different behavior.

In \cite{brethouwer_07}, a large-scale two-dimensional forcing is used, with grids up to $1024^2\times 320$ points. Computations are performed at high Reynolds number $Re$ and small Froude number $Fr$, varying the buoyancy Reynolds number ${\cal R}_B=ReFr^2$. Two regimes are identified, for low or high ${\cal R}_B$, with steep spectra and laminar layers in the former case, and the $k_{\perp}^{-5/3}$ spectra for kinetic and potential energy and turbulent layers in the latter case. These findings confirm previous works (see \cite{brethouwer_07} for a detailed review), and are often put in the context of atmospheric observations. Similarly, oceanic measurements of eddy diffusivity   have identified two regimes of mixing, in terms of the same parameter \cite{shih_05,ivey_08}. Using larger grid resolution and hyper-viscosity but similar forcing, it is shown in \cite{waite2011} that resolving or not the buoyancy scale may affect the outcome as far as energy distribution among Fourier modes is concerned, with steeper spectra when $L_B$ is well resolved, and that there is a sharp spectral break at the buoyancy scale as already predicted by \cite{weinstock_78}. Note that steep spectra mean that non-local interactions between widely separated modes are dominant. Moreover, when energy spectra are steeper than $k^{-2}$, dissipation takes place predominantly at large scale, and one cannot properly talk of an energy cascade phenomenon in the sense that dissipation acts over the entire spectrum. 

In \cite{kimura_12}, the choice is made of a cubic grid of $1024^3$ points, and the spectral data is also analyzed in terms of the wave-vortical decomposition introduced in \cite{craya_58, herring_74}. The spectra are found to be flat at large scale, a feature explained through the accumulation of sharp layers in the vertical direction. In \cite{almalkie_12}, a set of  large numerical simulations on grids of up to $4096^2\times 2048$ points are performed, and in these runs the Ozmidov scale is resolved. The horizontal spectra appear to follow again a $k_{\perp}^{-5/3}$ law, and it is noted that the direct cascade in the vertical direction provides a pathway to dissipation and is consistent with the generation of layers in the flow. This is also in agreement with the idea that the flow evolves towards the generation of layers such that the Froude number based on the vertical scale is of order unity \cite{billant_01}, a feature already observed empirically in \cite{metais2}. These results are confirmed by yet higher resolution runs \cite{debruynkops_13, debruynkops_14} on  grids of up to $8192^3$ points (in the homogeneous isotropic case), at unit Prandtl number and with buoyancy Reynolds numbers of up to 220. Such a high resolution allows for a detailed investigation of intermittency. Finally, in \cite{bartello_13}, the role of the buoyancy scale is confirmed; more importantly, the critical parameter to determine what scaling exponents prevail for the spectra seems to be the buoyancy Reynolds number ${\cal R}_B$: at large Reynolds number, the spectra are found to be independent of stratification.

What can be concluded from these past studies is that a consensus has not yet been reached as to whether there will be a universal description of such flows. In the present paper, we show that, as suggested already in \cite{cambon_94}, some of the ambiguities found in preceding studies may well be linked to a competition between several phenomena, namely on one hand the growth of slow modes with $k_{\perp} \approx 0$, and the dynamics of fast modes with $k_{\perp} \neq 0$ on the other hand.

\section{Methods} \label{sec:method}
\subsection{Equations}
The dynamics of a turbulent flow in a stably stratified environment can be described by the incompressible Navier-Stokes equations under the Boussinesq approximation. According to this model the three-dimensional velocity field ${\mathbf u({\bf x},t)}$ of components $(u,v,w)$, and the temperature fluctuations (or buoyancy field) $\theta({\bf x},t)$ obey the set of equations
 \begin{eqnarray} 
\partial_t {\mathbf u} +{\mathbf u} \cdot \nabla {\mathbf u}  &=&  -\nabla P - N \theta\  e_z + \nu \Delta {\mathbf u}+{\bf f}_V, \label{eq:mom} \\       
\partial _t \theta +{\mathbf u} \cdot \nabla \theta  &=& N w + \kappa \Delta \theta , \\  \label{eq:temp}
 \nabla \cdot {\bf u} &=&0;
\end{eqnarray}
\noindent where $P$ is the pressure, $N$ the Brunt-V\"ais\"al\"a frequency, $\nu$ the viscosity, ${\bf f}_V$ a velocity forcing term, and $\kappa$ the thermal diffusivity. As customary, the Brunt-V\"ais\"al\"a frequency is defined by $N=\sqrt{-(g/\theta ) (d\bar \theta /dz)}$, where $g$ is the gravitational acceleration and  $\bar \theta$ is a linear temperature profile. Equations (\ref{eq:mom}) to (\ref{eq:temp}) are solved with the pseudo-spectral Geophysical High-Order Suite for Turbulence (GHOST) code, which is parallelized with hybrid MPI/OpenMP programming, and has been tested on over 98,000 compute cores \cite{hybrid2011}. The code is based on a $2^{nd}$--order explicit Runge-Kutta temporal scheme, and uses a standard 2/3 de-aliasing rule in Fourier space.

The system can be characterized in terms of its energy, helicity, enstrophy, and dissipation rate, expressed either as a function of time or of Fourier space wavenumbers. The mean total energy $E_T$ is defined as the sum of the kinetic $E_V$ and potential $E_P$ energy 
$$ \frac{1}{2} \left<|{\bf u}|^2 + \theta\ ^2\right>=E_V+E_P =E_T, $$
and is a conserved quantity in the ideal limit. The brackets indicate the spatial mean. We also define helicity, the velocity-vorticity correlation, as
$$H_V = \left<{\mathbf u} \cdot {\mathbf \omega} \right>.$$
Helicity is an invariant of the inviscid non-stratified equations, it affects the cascade of energy in the presence of rotation \cite{helirot}, and has been observed to considerably slow-down the decay of turbulent energy in the presence of stratification \cite{rorai_13}.

The kinetic enstrophy, proportional to the kinetic energy dissipation, is given by $Z_V =  \left<\omega^2 \right>$. Similarly, the potential enstrophy is $Z_P =\left<|\nabla \theta | ^2 \right>$ and is associated with the dissipation of potential energy. As only the velocity field is forced, the total injection rate is simply given by
\begin{equation}
\varepsilon_V = \left< {\bf u} \cdot {\bf f}_V \right>. \label{epsilon}
\end{equation}
In the turbulent steady state, this quantity is expected to be equal (on the time average) to the total dissipation rate $\nu \left<\omega^2 \right> + \kappa \left<|\nabla \theta | ^2 \right>$.

\begin{figure} \centering
\resizebox{8.5cm}{!}{\includegraphics{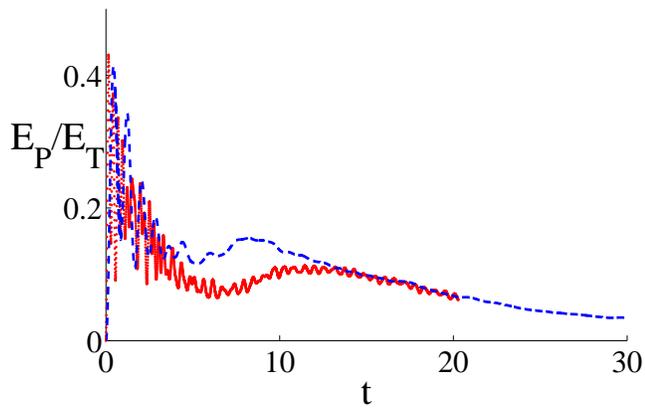}}
\caption{\label{PE_KE_t}
{\it (Color  online)} Temporal evolution of the ratio between  potential and total energy for the runs with $Fr\approx 0.1$ ($N=4$, dashed, blue line) and $Fr\approx 0.03$ ($N=12$, solid, red line). The oscillatory phase lasts longer for the more strongly stratified flow, but both ratios take comparable values after a transient.}
\end{figure}

Finally, we define the reduced energy and helicity spectra. In Fourier space the velocity autocorrelation function is noted $U_{ij}(k_x, k_y, k_z)$ and its trace is  $U({\bf k})$. Hence the axisymmetric kinetic energy spectrum is
 \begin{equation}
e_V(\mathbf{|k|}, \Theta)=\int U({\bf k}) |{\bf k}| \sin \Theta d\phi,
\end{equation}
where $\phi$ is the longitude with respect to the $k_x$ axis, and $\Theta$ is the co-latitude. By defining parallel $k_{\parallel}=k_z$, perpendicular $k_{\perp}=|{\bf k_{\perp}}|=|{\bf k}| \sin \Theta$, and isotropic $k=|\bf k|$ wavenumbers, we can calculate parallel, perpendicular, and isotropic reduced kinetic energy spectra as follows \cite{3072}
\begin{eqnarray}
E_V(k_{\parallel})&=&\int e_V(\mathbf{|k_{\perp}|}, k_{\parallel})dk_{\perp} \label{E2} , \\
E_V(k_{\perp})&=&\int e_V(\mathbf{|k_{\perp}|}, k_{\parallel})dk_{\parallel} \label{E1} , \\
E_V(k)&=&\int e_V(\mathbf{|k|},\Theta)|{\bf k}|d\Theta \label{E3} .
\label{aniso}\end{eqnarray}
Similar definitions hold for the potential and total energy, and for the helicity spectrum $h(|{\bf k}|,\Theta)$, which is related to the antisymmetric part of the velocity correlation tensor \cite{3072}. 

We can also distinguish between slow and fast mode spectra, namely:
\begin{eqnarray}
&E_{V, \textrm{slow}}&= e_V(\mathbf{|k_{\perp}|}=0, k_{\parallel}) \label{E4} , \\
&E_{V, \textrm{fast}}&= \int_{k_{\parallel}=0}^{k_{{\parallel}_{\max}}} \int_{|k_{\perp}|=1}^{|k_{\perp}|_{\max}} e_V(\mathbf{|k_{\perp}|}, k_{\parallel})dk_{\perp} dk_{\parallel} \label{E5}.
\label{aniso2}\end{eqnarray}
Equivalent definitions hold for $E_{P, \textrm{slow}}$, $E_{P, \textrm{fast}}$, $E_{T, \textrm{slow}}$, and $E_{T, \textrm{fast}}$. The slow modes satisfy the condition $\omega=0$, where $\omega$ is the frequency of gravity waves given by the dispersion relation $\omega=\sqrt{N^2k_{\perp}^2}/k$.  These modes correspond to ``pure'' eddies (vortical motions), and their characteristic time scale is the eddy turnover time. When $Fr<1$, the waves at large scales are faster than the eddies, or in other words, the wave period is faster than the turnover time. This is why the remaining modes, which correspond to a combination of eddies and waves, are often called ``fast'' modes.

The fluxes of kinetic and potential energy are respectively given by:
\begin{eqnarray}
&\Pi_{V}(k)&= \int_0^k \Gamma(k')dk' , \\
&\Pi_{P}(k)&= \int_0^k \mathrm{P}(k')dk' \label{pi},
\label{aniso2}\end{eqnarray}
with $\Gamma(k)={\bf u}^{\star}(k)\cdot\mathcal{F}({\bf u}\cdot\nabla {\bf u})_k$ and~$\mathrm{P}(k)={\bf u}^{\star}(k)\cdot\mathcal{F}(\rho\nabla \rho)_k$, where $\mathcal{F}$ denotes the Fourier transform. 

\begin{figure*}[h!tbp] \centering
\resizebox{8.5cm}{!}{\includegraphics{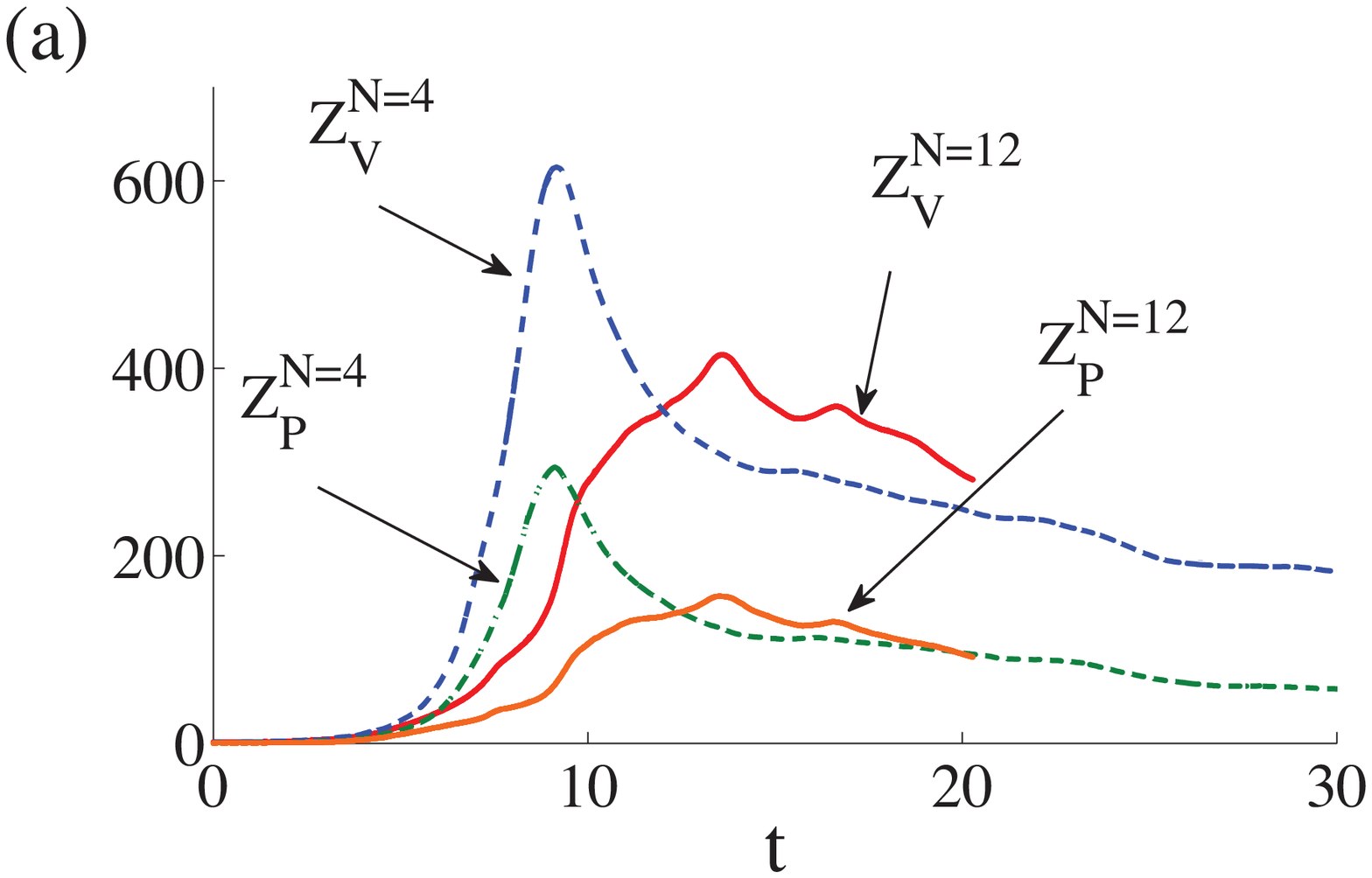}}
\resizebox{8.5cm}{!}{\includegraphics{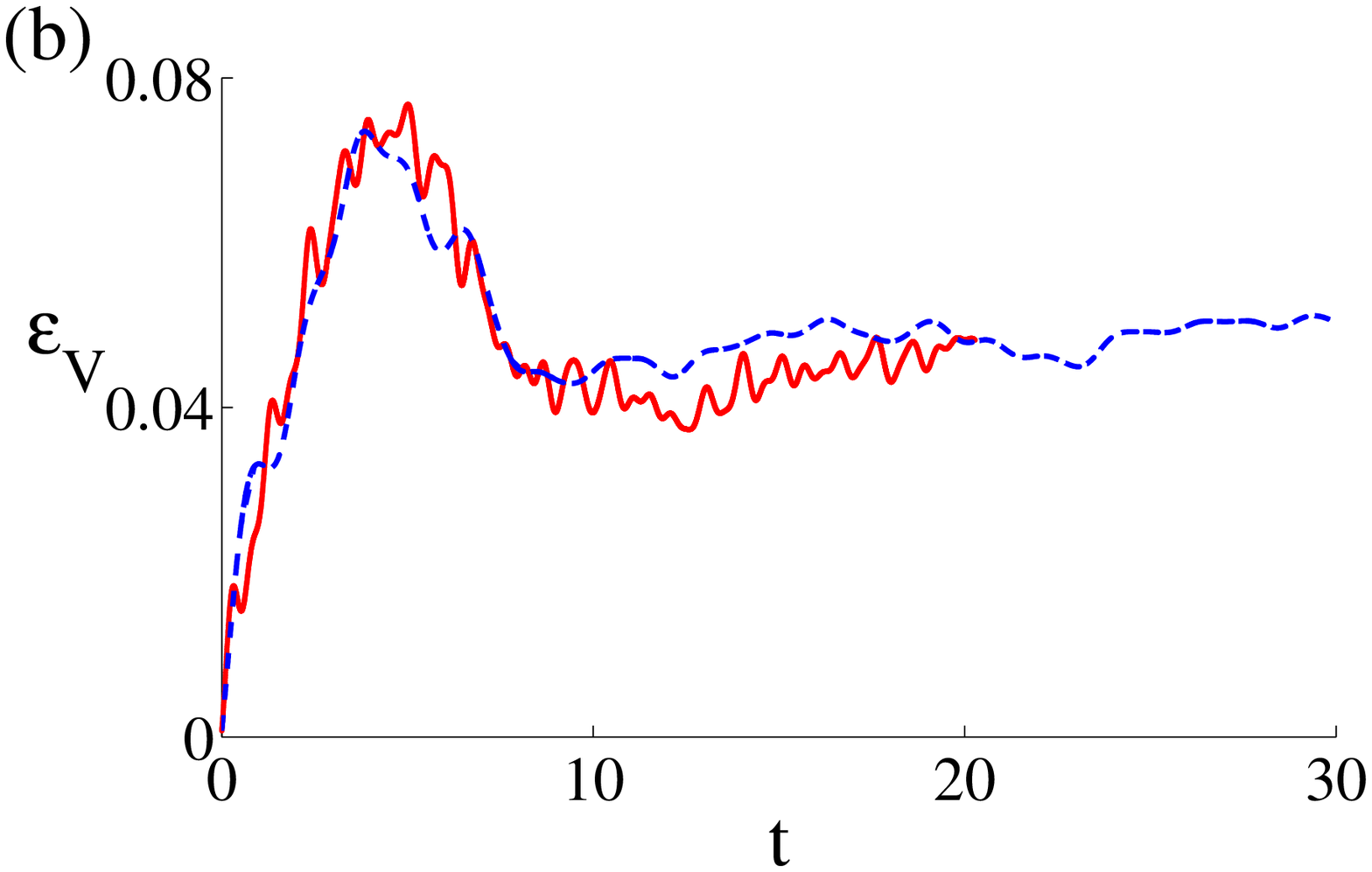}}
\caption{\label{eta_t}
{\it (Color  online)} Temporal evolution of (a) the kinetic enstrophy $Z_V=\left< |\omega|^2 \right>$ and the potential enstrophy $Z_P=\left<|\nabla \theta^2 |\right>$, and (b) the energy injection rate $\varepsilon_V=\left< {\bf u} \cdot {\bf f}_V \right>$  for the runs with $Fr\approx 0.1$ ($N=4$, dashed lines) and $Fr\approx 0.03$ ($N=12$, solid lines). More small scales are produced in the less stratified case at earlier times, but both flows evolve towards similar values by the end of the computations.}
\end{figure*} 

\subsection{Initial conditions and forcing}
Equations (\ref{eq:mom}) to (\ref{eq:temp}) are integrated numerically in a cubic domain of volume $V=(2\pi)^3$, discretized by an isotropic grid of 2048$^3$ points. The initial condition and the velocity forcing, ${\bf f}_V$, consist of randomly generated isotropic three-dimensional flows \cite{patterson} with injection wave number $k_F$ between $2$ and $3$. The forcing has amplitude  $f_\textrm{rms}=0.22$, chosen to yield an  approximately unitary r.m.s.~velocity ($u_\textrm{rms}=1$) in the turbulent steady state. We impose $\nu=10^{-4}$, which guarantees that the Kolmogorov scale for a homogeneous isotropic flow with the same parameters and discretization is well resolved \cite{mininni_nonlocal}. Note this is a conservative choice, since in wave turbulence the energy spectrum is expected to be steeper, and therefore the small scales are expected to be less energetic. As a result of these choices, the ratio between the smallest and largest scales resolved in our calculations is about $\approx 700$, and the Reynolds number is $Re\approx 25000$.

We perform two runs with different values of the Brunt-V\"ais\"al\"a frequency, resulting in $Fr\approx 0.1$ (for $N=4$), and $Fr\approx 0.03$ (for $N=12$). The buoyancy Reynolds number ${\cal R}_B=ReFr^2$ is correspondingly ${\cal R}_B=250$ and ${\cal R}_B=27$. The calculations are carried out for respectively 30 and 20 time units. Finally, in both runs we consider a unitary Prandtl number $Pr=\nu/\kappa=1$.

\subsection{Characteristic scales}
For our simulations of stratified flows, the relevant length scales are:

\begin{enumerate}
\item[(i)] The overall size of the periodic domain, equal to $L_0=2\pi$ in dimensionless units, and with associated wavenumber $k_0 = 2\pi /L_0$.
\item[(ii)] The scale at which energy is injected into the system, $L_F=2\pi/k_F$.
\item[(iii)] The buoyancy scale, $L_B=2\pi / k_B$, with $k_B = N/u_\textrm{rms}$, characteristic of the vertical shear.
\item[(iv)] The scale at which isotropy (and presumably a Kolmogorov energy spectrum) is recovered, namely the Ozmidov scale $\ell_\textrm{oz}=2\pi/k_\textrm{oz}$, with $k_\textrm{oz} = (N^3/\varepsilon)^{1/2}$.
\item[(v)] The dissipation scale, $\eta=2\pi/k_\eta$, with $k_\eta = (\varepsilon/\nu^3)^{1/4}$.
\item[(vi)] The smallest scale resolved in the DNS, namely $\ell_\textrm{min}=2\pi/k_\textrm{max}$. Because of the Fourier transform the pseudospectral code is based upon, and the 2/3-rule for removing aliasing $k_\textrm{max}=n/3 \approx 700$ where $n$ is the number of grid points per dimension.
\end{enumerate}
In Table \ref{scales} we report the values of the wavenumbers associated with these characteristic scales. 

The Ozmidov and the dissipation scales are usually evaluated by estimating $\varepsilon \approx u_\textrm{rms}^3/L_F$. However, this estimation is valid for isotropic and homogeneous turbulence, while in a stratified flow the energy injection rate, flux, and dissipation rate can be strongly affected by the waves. We estimate then $\varepsilon \approx \varepsilon_V$, following the definition in Eq.~(\ref{epsilon}) which corresponds to the effective rate of transfer in the flow. Using $\varepsilon_V$ yields a value one order of magnitude smaller for the injection and dissipation rates (see Sec.~\ref{sec:temporal}). The estimates obtained in this way are marked by a star in Table \ref{scales}: $k_{oz}^*$ and $k_{\eta}^*$. As will be shown later, these quantities give a better estimation of at what scales transitions occur in the flow. In practice, a well-resolved run requires $k^*_\eta<k_\textrm{max}$ as dissipation starts to dominate the dynamics at this wavenumber; observe that this is satisfied by our simulations. Also, it can be easily shown that the Ozmidov scale is resolved (e.g., $k^*_\textrm{oz} < k^*_\eta$), when ${\cal R}_B=ReFr^2 \ge 1$; at the scale at which $k^*_\textrm{oz} = k^*_\eta$ the wave period equals the eddy turnover time.

A second effective estimate of the vertical characteristic scale, which can be associated with the bouyancy scale, is given by the integral scale based on the parallel potential energy spectrum, as layers tend to develop more clearly in the temperature:
\begin{equation}
L_{B}^*=2\pi\frac{\int E_P(k_{\parallel})/k_\parallel dk_{||}}{\int E_P(k_{\parallel}) dk_{||}} .
\end{equation}
The corresponding wavenumber $k_B^*=2 \pi /L_{B}^*=2$ is also reported in Table \ref{scales}.

\begin{table}
\begin{ruledtabular}
\begin{tabular}{lcc}
Runs  & $N=4$ & $N=12$ \\
\hline
\hline
$k_0$ & 1 &  1  \\
$k_F$ & 2-3 & 2-3 \\
$k_\textrm{max}$ & 683 & 683\\
\hline
$k_B$    & 4 & 12 \\
$k_B^*$ & 7 & 8 \\
\hline
$k_\textrm{oz}$     & 13 & 66\\
$k_\textrm{oz}^*$  & 36 & 186 \\
\hline
$k_\eta$     &  795 & 795 \\
$k_\eta^*$  & 472 &  472 \\
\end{tabular}
\end{ruledtabular}
\caption{Wavenumbers corresponding to the box size ($k_0$), injection scale ($k_F$), grid resolution ($k_\textrm{max}$), bouyancy scale ($k_B$ and $k_B^*$, where wavenumbers without a star are computed using $\epsilon \approx u_\textrm{rms}^3/L_F$, and wavenumbers with stars are computed using the 
 measured
injection rate $\epsilon_V$), Ozmidov scale ($k_\textrm{oz}$ and $k_\textrm{oz}^*$), and dissipation scale ($k_\eta$ and $k_\eta^*$).}
\label{scales} \end{table}

Finally, we attempt to assign physical values, characteristic of the atmosphere and the oceans, to the run at the smallest Froude number, with $Fr=0.03$ and $Re \approx 2.5 \times 10^4$. For the atmosphere we assume $u_\textrm{rms}=1$ ms$^{-1}$ and $L_0=1000$ m (roughly the size of a small convective cell). Hence it is readily found that $N \approx 3.3 \times 10^{-2}$ s$^{-1}$, $\varepsilon_V \approx 4\times 10^{-5} $m$^2$s$^{-3}$ (per unit mass), and $\nu=0.04$ m$^2$s$^{-1}$, clearly too large for the atmosphere, as expected, given the limited grid resolution. However, the energy injection rate is close to atmospheric values, which yield $\varepsilon_V \approx 10^{-6}$--$10^{-5} $m$^2$s$^{-3}$ from data analysis of aircraft measurements \cite{Lindborg99} and of satellite images \cite{Heas12}. From these values it also follows that $L_B  \approx 190$ m and $\ell_\textrm{oz}  \approx  33$ m, to be compared to the Kolmogorov dissipation scale of $ \approx 3$ m and to the grid resolution of $\ell_\textrm{min}  \approx 1.4$ m. If we consider the ocean instead, the typical velocity is ten times smaller. Hence, given the same Reynolds and Froude numbers,  $L_B$ and $\ell_\textrm{oz}$ remain the same, while the Brunt-V\"ais\"al\"a frequency and the viscosity are reduced by an order of magnitude. Yet, $\nu=0.004$ m$^2$s$^{-1}$ is three orders of magnitude larger than realistic values.

\begin{figure*}[h!tbp] \centering
\resizebox{8cm}{!}{\includegraphics{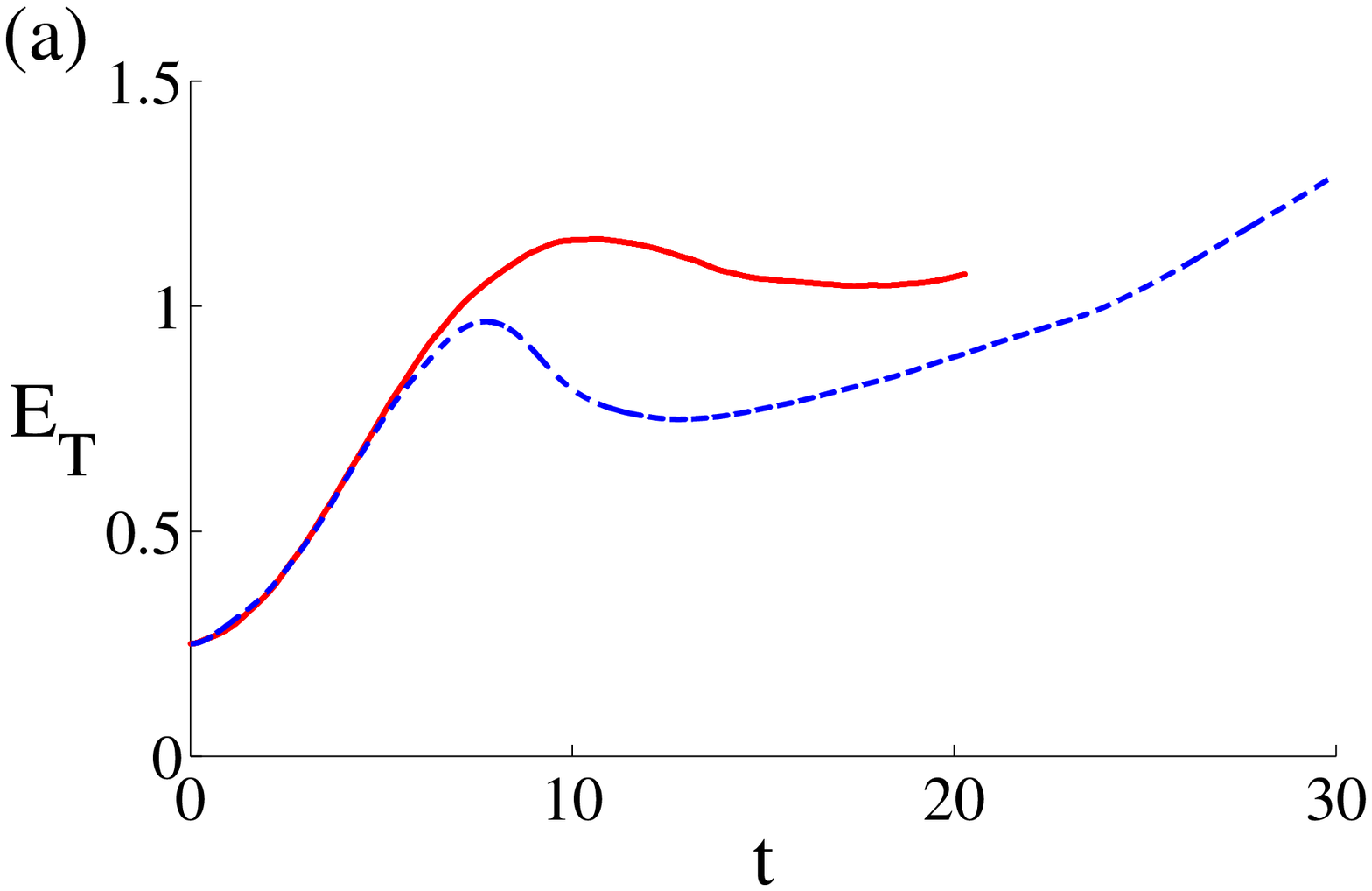}}
\resizebox{8cm}{!}{\includegraphics{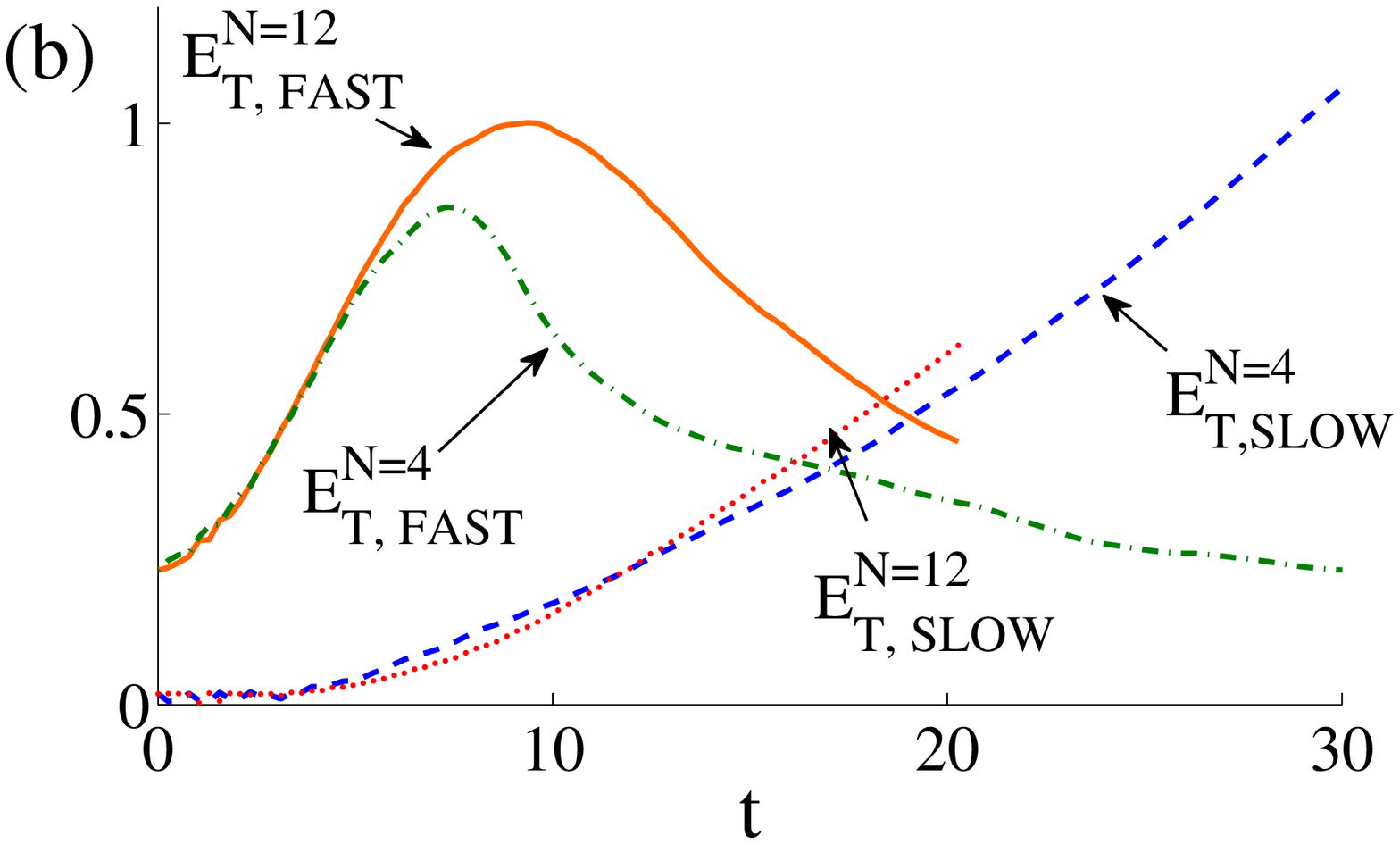}}
\caption{\label{L_t}
{\it (Color  online)} (a) Temporal evolution of the total energy for the runs with $Fr\approx 0.1$ ($N=4$, dashed, blue line) and with $Fr\approx 0.03$ ($N=12$, solid, red line).  (b) Temporal evolution of the energy in the fast and slow modes for the same two runs (see label on the curves). Note the dominance of slow modes as time evolves.
} \end{figure*} 

\section{Results} \label{sec:numerics}
We present first in Sec.~\ref{sec:temporal} the temporal behavior of small and large scale quantities, integrated over the entire domain. Quantities associated with the dynamics of small scales ($Z_V$, $Z_P$ and $\varepsilon_V$) reach a steady turbulent regime at an early stage. In contrast, quantities associated with the energetics of the large scales ($E_T$, $E_{T, fast}$ and $E_{T, slow}$) do not converge to a statistical steady state by the end of our calculations. In this case, stationarity is primarily prevented by the monotonic growth of slow modes, as also found for example in \cite{smith_02}. Then, in Sec.~\ref{sec:spectra} we present the energy and helicity distribution among Fourier modes, including a study of spectral anisotropy. Energy and helicity spectra are averaged in time to obtain a representative statistical behavior, and early times are excluded from the average as the turbulent regime is established only after an initial transient in which the flow adapts to the forcing and develops small-scale structures. Finally, in Sec.~\ref{sec:fluxes} we comment on the energy fluxes.

\subsection{Temporal evolution of global quantities\label{sec:temporal}}
Figure \ref{PE_KE_t} shows the ratio of the potential to the total energy as a function of time. The eddy turn-over time is $\tau_{NL}\approx 2.5 t$. After an initial transient, both curves, independently of the stratification strength, reach a value of  $E_P/E_T\approx 0.1$ about $t\approx14$, followed by a slow monotonic decrease. This ratio is comparable with that found in \cite{brethouwer_07, kimura_12} for  similar values of ${\cal R}_B$.

In Fig.~\ref{eta_t}(a) the temporal evolution of  the kinetic and potential enstrophy is shown. The initial transient is characterized by the development of small-scales through non-linear mode coupling. This mechanism is less efficient at low $Fr$ as testified by the smaller values of the two enstrophies at early times, resulting also in a smaller value for the dissipation of total energy $\left< \nu |\omega|^2  + \kappa |\nabla \theta|^2 \right>$ for the $N=12$ run. The enstrophy maxima occur earlier in terms of the eddy turn-over time, but not in terms of the buoyancy period, for the less stratified flow. Turbulence can be said fully developed beyond the peak of enstrophy where dissipation reaches the smallest scales. Interestingly, at later times, the curves of the two runs merge and undergo a slow decay. This behavior can be identified with the achievement of a turbulent steady state, at least at small scales and as long as the slow modes are not dominant. Note that as in the case of the ratio $E_P/E_T$, the enstrophies (and as a result, the energy dissipation rates) also tend to converge to similar values independently of the two stratification strengths considered.

As a comparison, in Fig.~\ref{eta_t}(b) we show the temporal behavior of the energy injection rate computed using Eq.~(\ref{epsilon}). The similarity of the two runs is remarkable. A dimensional Kolmogorov-like evaluation of the energy injection rate for fully developed turbulence $\varepsilon \approx u_{rms}^3/L_F$, yields, for our r.m.s.~velocity and forcing scale,  $\varepsilon \approx 0.4$, an estimate one order of magnitude larger than the numerical value $\varepsilon_{V}\approx 0.04$. This latter value is also compatible with the dissipation rate at late times obtained from $\left< \nu |\omega|^2  + \kappa |\nabla \theta|^2 \right>$. The order of magnitude difference between the Kolmogorov-like estimation and the actual values of injection and dissipation can be understood as in wave turbulence the energy transfer rate is expected to be smaller than $\approx u_{rms}^3/L_F$ by a factor $Fr$, as indicated by numerous studies \cite{newell_01, nazar}. However, it should be noted that this argument fails to explain why $\varepsilon_{V}$ in the two simulations has similar values independently of the value of $Fr$.
One possibility is that much of the dissipation occurs in the strong gradients that develop in the vertical in order to insure that the Froude number based on a characteristic vertical scale is of order unity, in which case the weak turbulence argument can only apply to the horizontal dynamics. 
In other words, the lesser dissipation in the horizontal for smaller Froude number is compensated almost exactly by the increased dissipation in the vertical. This phenomenon is related to the distribution of energy between the potential and kinetic modes on the one hand, and between the vertical and horizontal kinetic energy on the other hand.

We then conclude, from Fig.~\ref{eta_t}, that small scales have saturated and we remark that the two runs have a tendency towards identical dissipation, a surprising result since the buoyancy Reynolds numbers differ by almost an order of magnitude (but are in both cases above a critical ${\cal R}_B\approx 10$).
This result is consistent with the recent finding in \cite{bartello_13} that the energy spectra are independent of stratification at sufficiently high Reynolds number. However, note that while this is indeed the case for the spectrum of small scale fluctuations, it is not the case at large scales as will emerge from the analysis of the spectra.
 
\begin{figure*}[h!tbp] \centering
\resizebox{8cm}{!}{\includegraphics{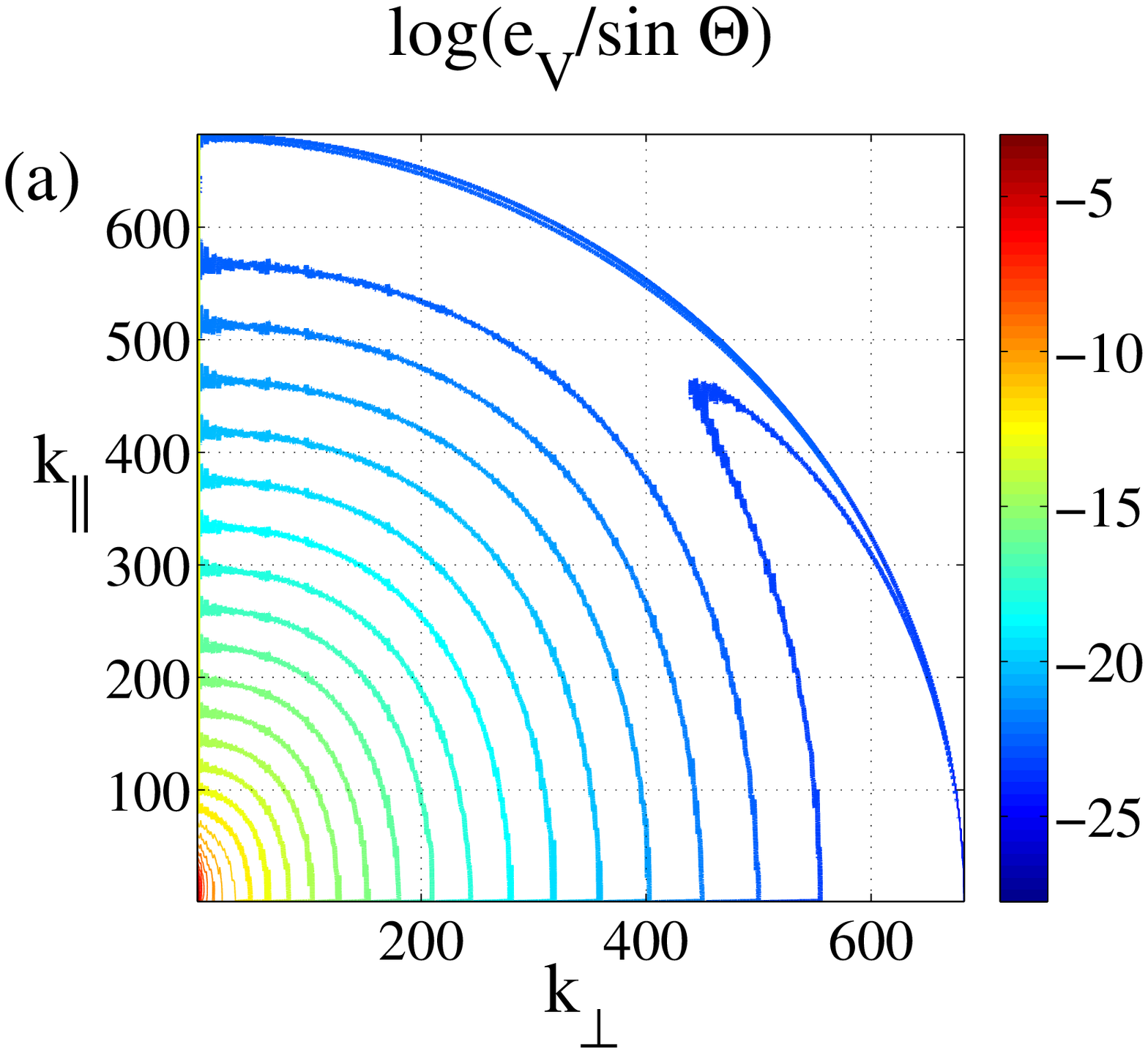}}
\resizebox{8cm}{!}{\includegraphics{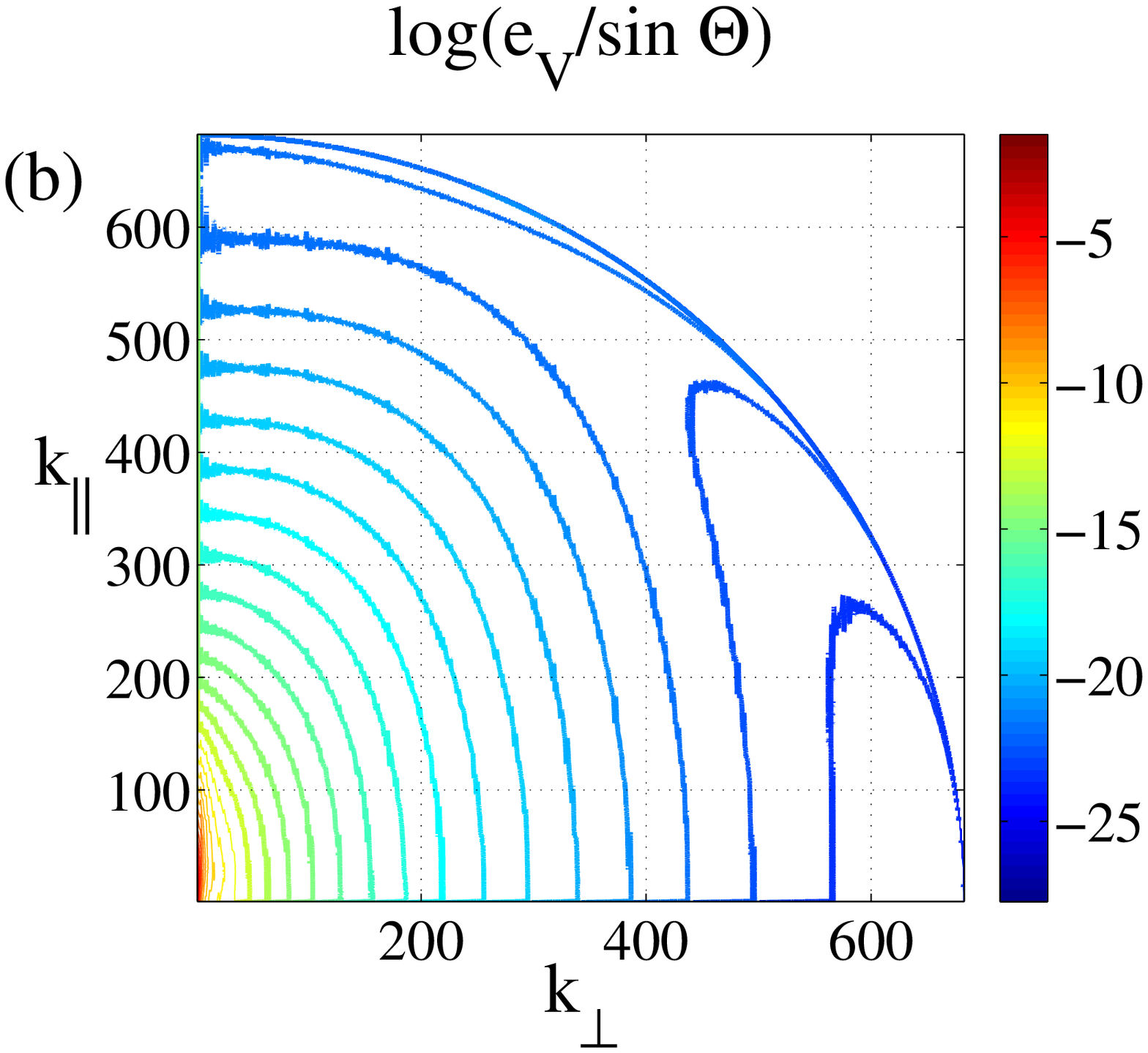}}
\caption{\label{cont}
{\it (Color  online)} Contours of the two-dimensional (axisymmetric) kinetic energy spectrum, averaged over time for the runs with (a)  $Fr\approx 0.1$ ($N=4$), and (b) $Fr\approx 0.03$ ($N=12$). Note how energy tends to accumulate in slow modes (modes with $k_\perp \approx 0$, also associated with vertically sheared horizontal winds) with elongated isocontours specially at large scales. Isocountours in (b) are more vertically elongated, indicating stronger anisotropy.}
\end{figure*} 

As a first indication of differences at large scales, the time evolution of the total energy is shown in Fig.~\ref{L_t}(a). As turbulence develops, the total energy grows until it reaches a peak. At a later stage a fluctuating behavior, characteristic of the turbulent steady state, is expected. However, after the peak, we observe a new monotonic increase with a timescale which is larger than characteristic times such as $1/N$ or $\tau_{NL}$. It is evident from Fig.~\ref{L_t}(b) that the energy increase is due to the growth of the slow modes, while the energy in fast modes gradually decreases. We verified on a lower-resolution run and with forcing at substantially smaller scale that the growth of the fast modes saturates after thirty turn-over times, and a steady state is indeed reached \cite{marino_14}.

Several remarks follow from considering the dispersion relation of internal gravity waves
$$\omega= \pm k^{-1}\sqrt{N^2k^2_{\perp}} \ . $$
A review of various theoretical approaches to stratified turbulence viewed as a superposition of  internal waves can be found for example in \cite{muller_86, staquet_rev_02, sagaut_cambon_08, polzin}. Here it is of interest to recall that three-wave interactions at resonance play a central role in closing the cumulant expansion; in the present case, for a usual triad of modes ${\bf k}$, ${\bf p}$ and ${\bf q}$ satisfying ${\bf k} = {\bf p} + {\bf q}$ , they read
$$s_k \frac{k_{\perp}}{k} = s_p \frac{p_{\perp}}{p} + s_q \frac{q_{\perp}}{q} ,$$
where $s_k$, $s_q$ and $s_q=\pm 1$ depending on the branch of the dispersion
 relation used. As remarked in \cite{waleffe_93, smith_99}, this resonance condition is readily satisfied for $k_{\perp} \approx p_{\perp} \approx q_{\perp} \approx 0$. Hence, it can be inferred that the wave-wave interactions lead to a build-up of energy at $k_{\perp} \approx 0$, i.e., in the so-called slow modes. The disctintion between energy in fast modes and in slow modes in Fig.~\ref{eta_t} is compatible with this build-up of energy in modes with $k_{\perp} \approx 0$.

The accumulation of energy for $k_{\perp}=0$ was already noticed in \cite{cambon_94}, using a two-point closure of turbulence, the so-called EDQNM2 (Eddy Damped Quasi-Normal Markovian 2 closure). It was also found in \cite{smith_02, brethouwer_07} in DNS, and attributed to the growth of the so-called vertically sheared horizontal winds. It is to be noted that such winds, which are off-diagonal elements of the velocity gradient matrix, constitute 
 the vorticity field for negligible vertical velocity and as such they make the velocity field non potential. In the presence of rotation, these are the so-called thermal winds which are a-geostrophic corrections to geostrophic balance. It is interesting (and may be viewed as somewhat paradoxical) that resonant interactions of gravity waves can lead to the growth of vortical modes which eventually come to dominate the flow.

\begin{figure*}[h!tbp] \centering
\resizebox{8cm}{!}{\includegraphics{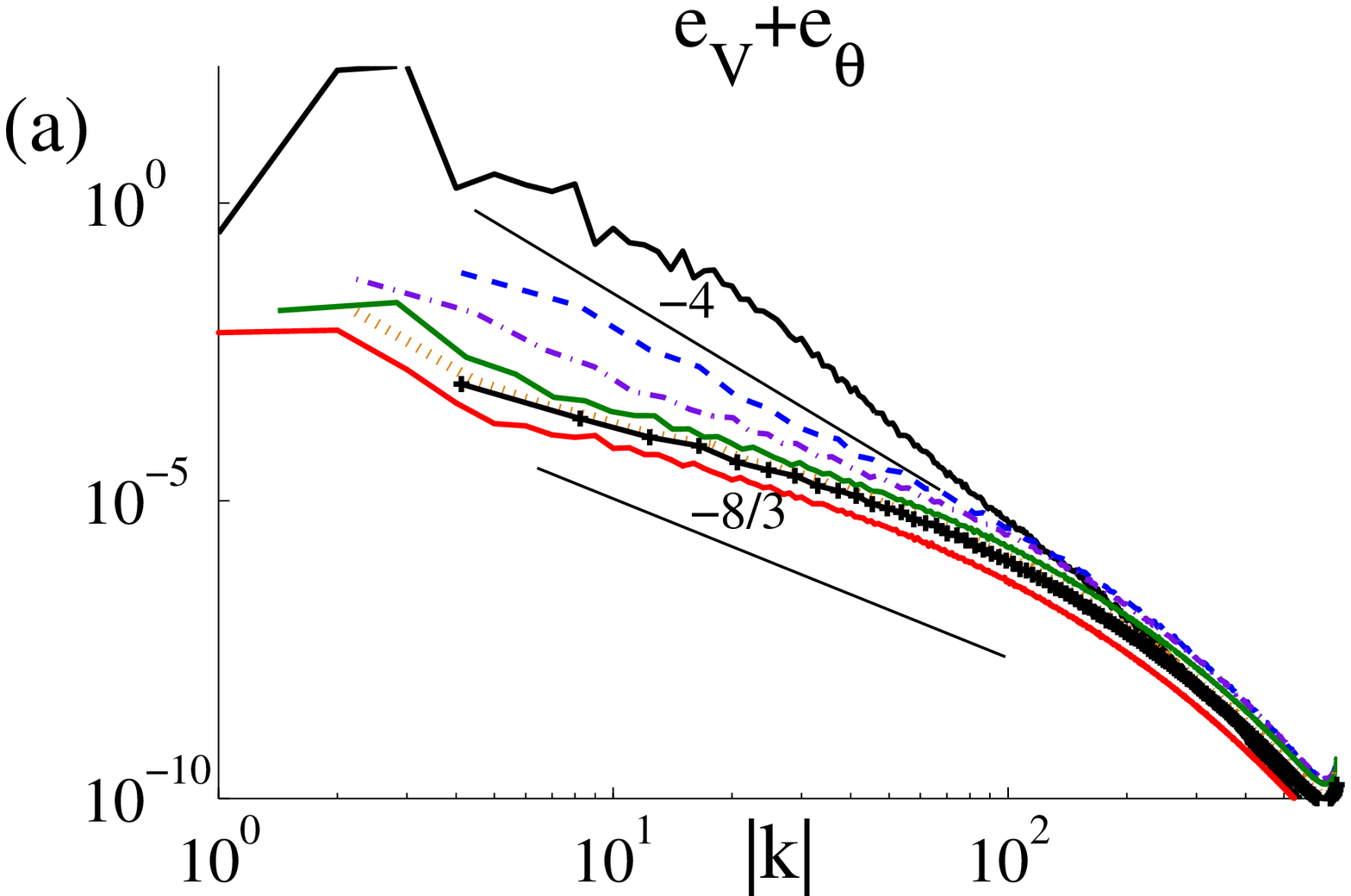}}
\resizebox{8cm}{!}{\includegraphics{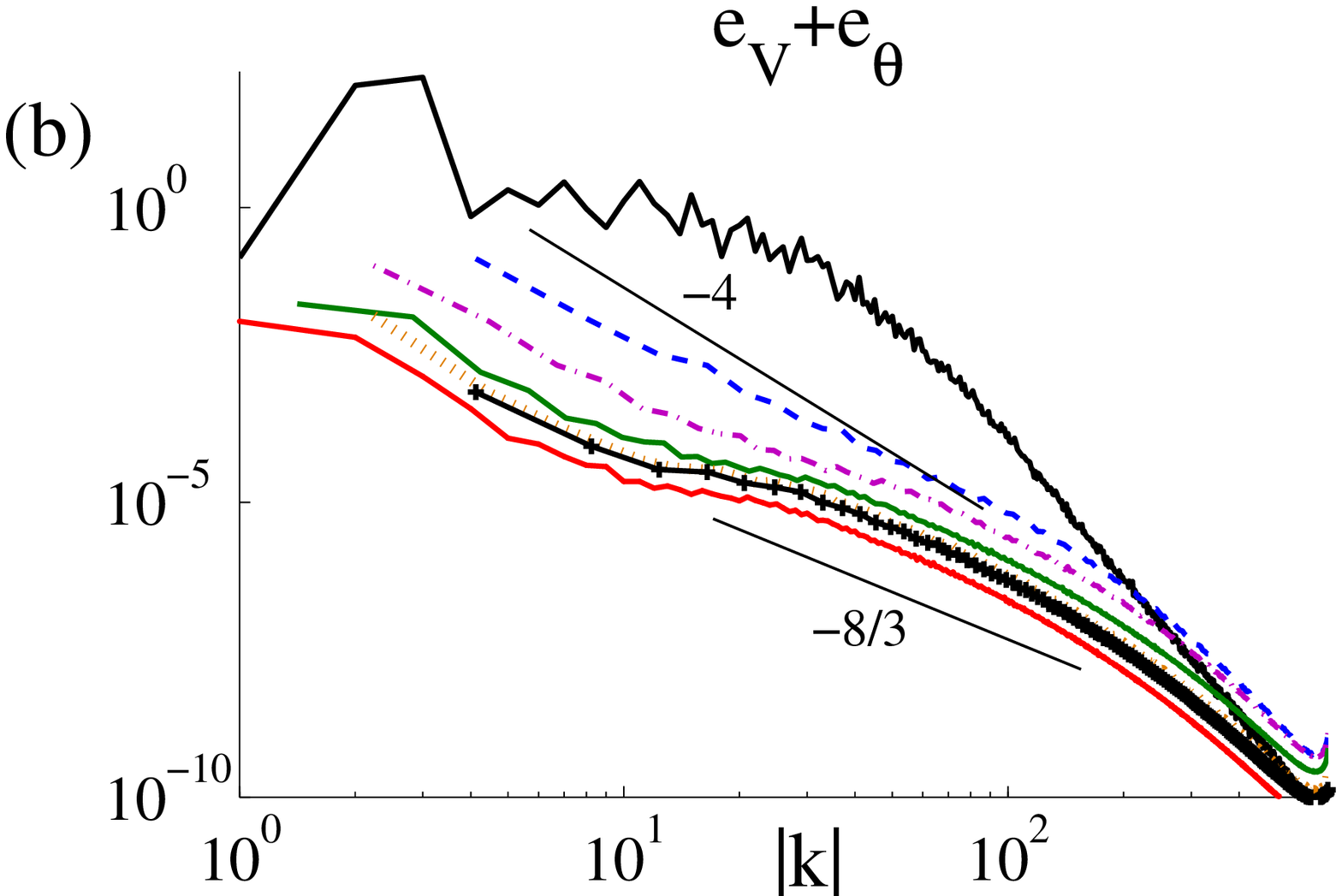}}
\caption{\label{2Dspec}
{\it (Color  online)} Two-dimensional (axisymmetric) total energy spectra for $Fr\approx 0.1$ ($N=4$, left) and $Fr\approx 0.03$ ($N=12$, right) for different co-latitudes (with respect to the vertical axis $k_\parallel$ in Fourier space): 
$\Theta=0$ (black, solid line),
$\Theta\approx 14^\circ$ (blue, dashed line),
$\Theta\approx 26^\circ$ (purple, dash-dotted line),
$\Theta=45^\circ$ (green, dash-triple-dotted line),
$\Theta\approx 64^\circ$ (yellow, dotted line),
$\Theta\approx 76^\circ$ (black, crossed line),
and finally $\Theta=90^\circ$ (red, solid line).
Observe the dominance of energy in the $k_{\perp}=0$ slow modes ($\Theta=0$), all the way to the Ozmidov scale where isotropy starts to recover. Solid lines indicate power laws as a reference. 
}\end{figure*} 

We finally comment on the temporal dynamics of helicity and relative helicity, $\sigma_V=H_V (E_VZ_V)^{-1/2}$. Although we do not use a helical forcing, we are not imposing the forcing to be completely non-helical. As a result, there is a small amount of helicity in the flow at late times. The time evolution is at follows: helicity starts from zero and it undergoes an oscillatory transient, longer for stronger  stratification. In both runs, $\sigma_V$ then grows slowly in time until it reaches a final value of $\approx 0.12$ for the $N=4$ run. Similarly, and as will be shown later, the relative spectral density $H_V(k)/[kE(k)]$ remains low, of the order of $0.05$, except in the vicinity of the forcing scales. These values are too small to affect the energetics of the flows for either runs. In \cite{rorai_13}, the dynamics of helicity in freely decaying stratified turbulence was examined in terms of a possible balance between its production and dissipation. It was found that only when the initial condition was an ABC (maximally helical) flow, the energy spectrum was modified by the slowed-down dynamics inherent to the helical case, a situation also found in rotating flows \cite{teitelbaum}. 
 
\begin{figure*}[h!tbp] \centering
\resizebox{10cm}{!}{\includegraphics{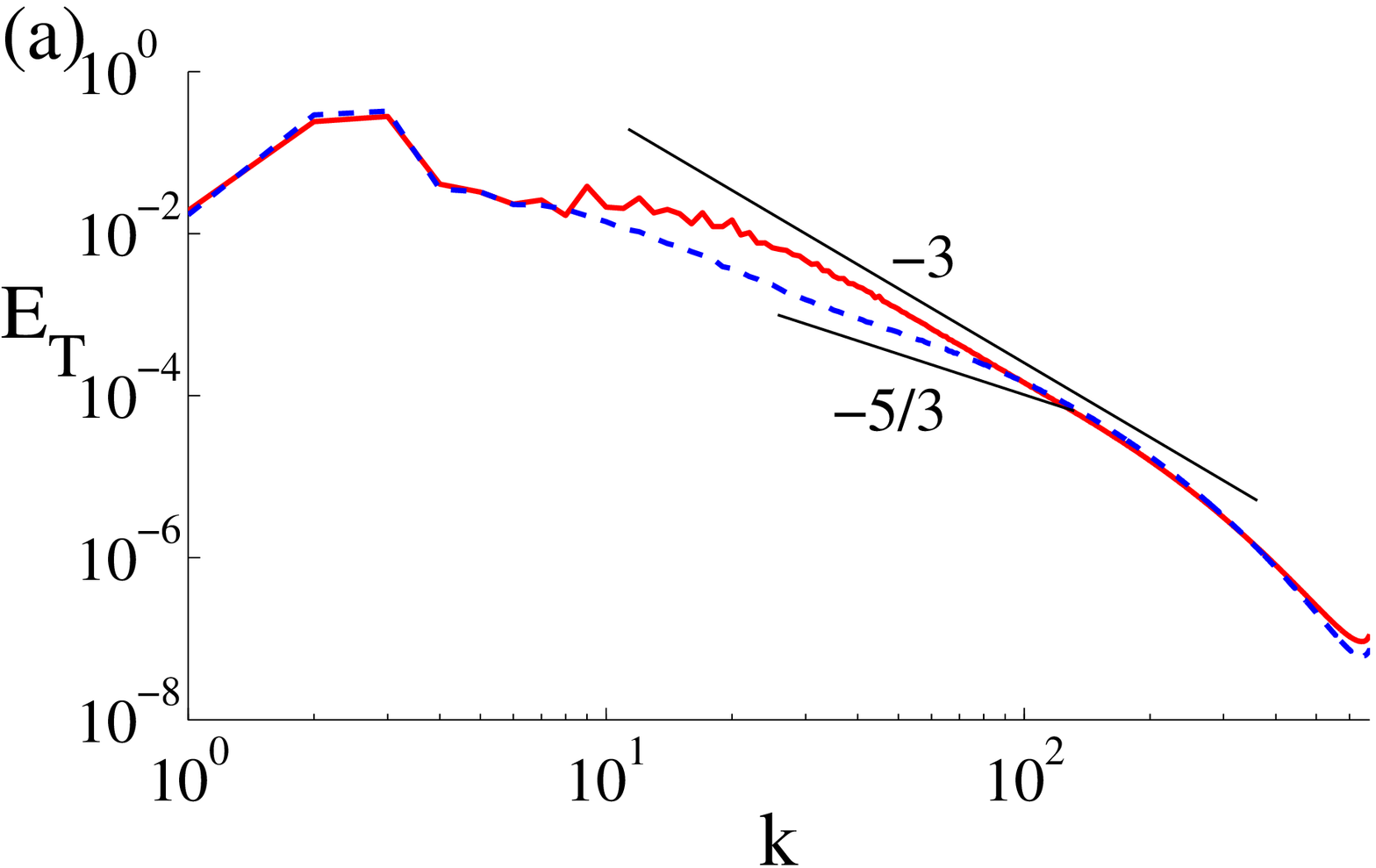}}
\resizebox{18cm}{!}{\includegraphics{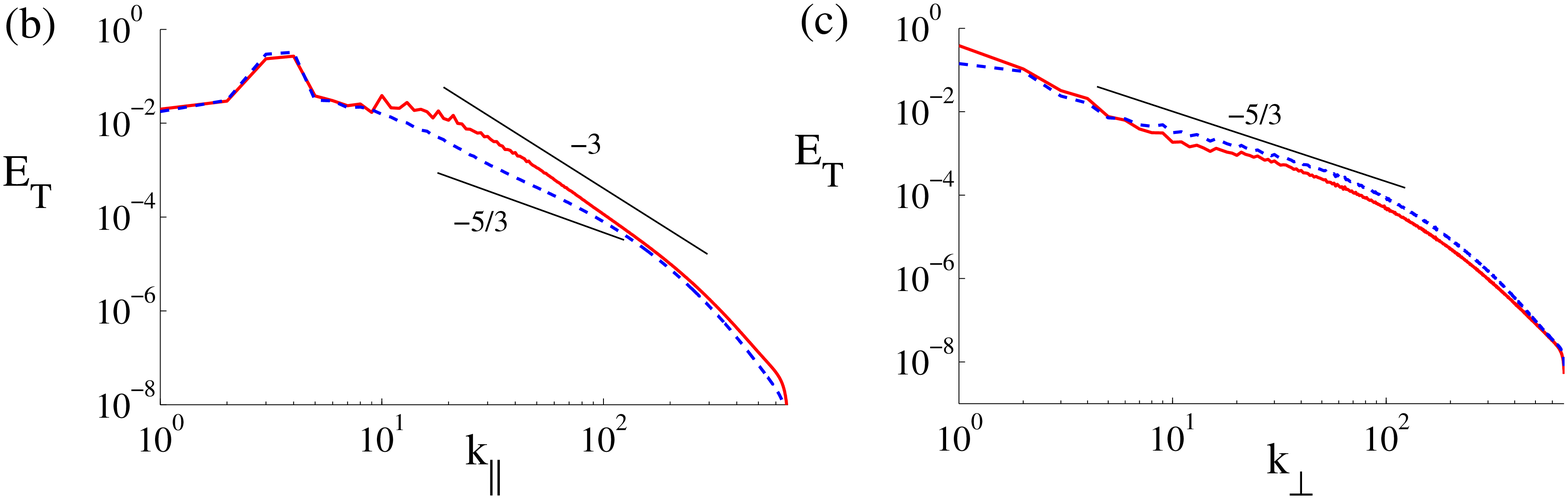}}
\resizebox{18cm}{!}{\includegraphics{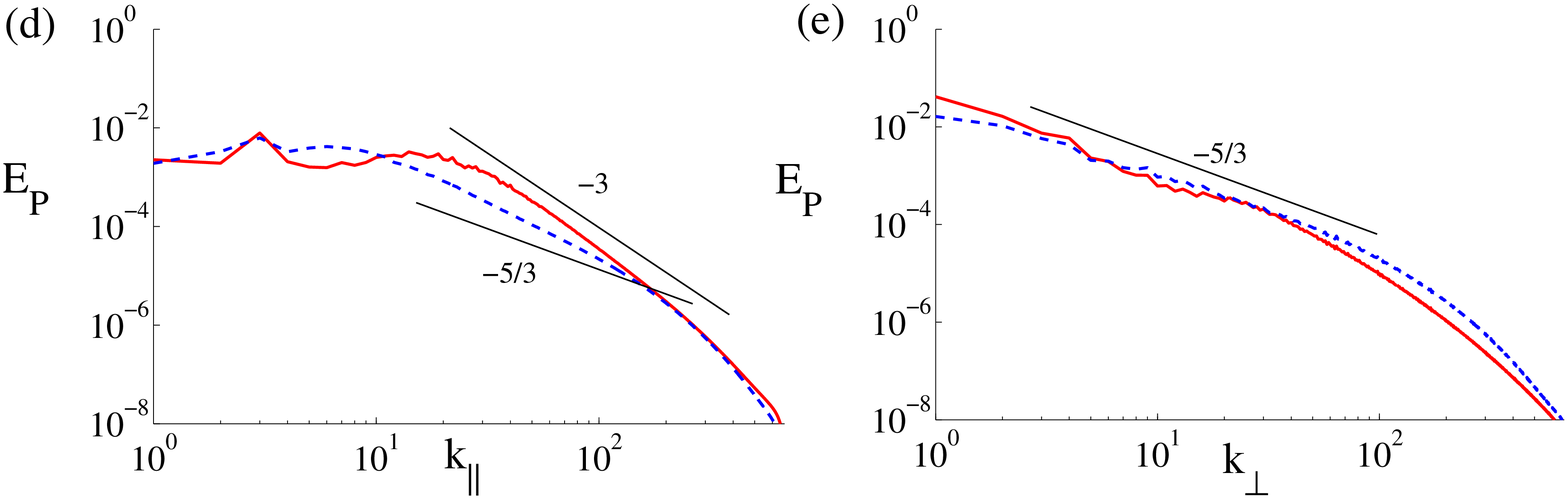}}
\caption{\label{Ek}
{\it (Color  online)} Total energy spectra as a function of (a) isotropic wave number, (b) parallel wave number, and (c) perpendicular wave number. The potential energy spectra are shown also as a function of (d) parallel wave numbers, and (e) perpendicular wave numbers. Solid (red) lines correspond to the run with $Fr\approx 0.03$ ($N=12$ run), while dashed
 (blue)
 lines correspond to the run with $Fr\approx 0.1$ ($N=4$). Solid straight lines indicate some power laws as a reference. The buoyancy scale is identifiable as a break in the potential energy spectrum as a function of $k_{\parallel}$.
} \end{figure*}

\subsection{Energy distribution among modes\label{sec:spectra}}
\subsubsection{The development of anisotropy}
We now consider the energy and helicity spectra. We first show, in Fig.~\ref{cont}, the isocontours of the axisymmetric kinetic energy spectrum $e_V({k_{\perp},k_{\parallel}})$ (normalized by $\sin \Theta$ to obtain circular isocontours in the case of an isotropic flow). When stratification is stronger, the flow is more anisotropic, i.e., the contours are more stretched along the vertical axis. In other words, there is more energy in the slow modes with $k_\perp \approx 0$, as described before. However, the anisotropy is not the same for all wavenumbers; small wavenumbers (large scales) tend to be more anisotropic than large wavenumbers (small scales). This is more clear in the less stratified flow. Indeed, a rough estimate of the recovery of isotropy can be made by considering the wavenumber for which the contour lines approach a circular shape. This approximately occurs at large wavenumbers for $N=4$, whereas for $N=12$ isotropy is only barely recovered at the smallest resolved scales. 

The axisymmetric total energy spectra $e(k_{\perp}, k_{\parallel})=e_V(k_{\perp}, k_{\parallel})+ e_P(k_{\perp}, k_{\parallel})$ for various values of the co-latitude $\Theta$ are shown in Fig.~\ref{2Dspec}. The dominance of the slow modes (i.e., modes with $k_{\perp}=0$) is evident for both runs. However, the modes with $k_{\perp}=0$ have different spectra at large scales depending on the value of $N$. While in the run with $N=4$ the spectrum is steep, in the run with $N=12$ the spectrum is shallower and almost flat (see, e.g., \cite{kimura_12} for previous observations of the flat spectrum). We recall that the isotropic spectra for a given $k$ is obtained by summing over $\Theta$ [see Eq.~(\ref{E3})]. Since the energy distribution depends on the stratification intensity, the isotropic spectrum for stratified fluids will be different for different stratifications. Moreover, some of the spectra in Fig.~\ref{2Dspec} are rather steep [note that a spectrum $\sim k^{-4}$ in $e(k_{\perp}, k_{\parallel})$ corresponds, after integration, to a power law $\sim k^{-3}$ in the isotropic spectrum]. For flows whose isotropic spectrum has an inertial index $\alpha$ which falls outside the range $-1<\alpha<-3$, arguments for locality of interactions do not hold, and interactions between modes can become non-local. Both these effects can give rise to non-universality of the spectrum (i.e., as a result of the existence of different physical regimes,  or as the result of non-locality). This has been discussed often in the context of numerous and detailed observations of oceanic and atmospheric flows \cite{polzin}, and also noted for example in the  framework of internal waves in the ocean in  the hydrostatic (and irrotational) limit \cite{lvov_10}.

As already mentioned, in the weakly stratified run, the dominant modes seem to follow a $k_{\parallel}^{-4}$ spectrum, the so-called saturated spectrum (in these units, the isotropic spectra being recovered after one integration over wavenumber), whereas a short range of wavenumbers at small scales is compatible with a $k_{\perp}^{-8/3}$ law (corresponding to a Kolmogorov spectrum after integration). Moreover, note that $k_\textrm{oz}^* \approx 36$ in this run; for smaller wavenumbers all angular spectra start to collapse indicating a return to isotropy. On the other hand, in the run with $N=12$, $k_\textrm{oz}^* \approx 186$ and the spectra only collapse in the dissipative range. It is interesting 
and significant
to note that the wavenumbers based on the 
 dynamical
injection rate obtained from $\epsilon_V$ ($k_\textrm{oz}^*$) give a much better estimation of the return of isotropy than those obtained from the Kolmogorov estimate $\epsilon \sim u_\textrm{rms}^3/L_F$ ($k_\textrm{oz}$).

In Ref.~\cite{chung_12} it is suggested to analyze the data once the slow modes, corresponding to the $k_{\perp}=0$ modes, are removed. The analysis of the flow in terms of two-dimensional spectra allows for such a reduction, since the $k_{\perp}=0$ modes are confined to the $\Theta=0$ angle (black solid line in Fig.~\ref{2Dspec}). Remarkably, the data seems to separate into two ranges of scales with different behaviors, one range similar to the $\Theta=0$ spectrum, and another similar to the $\Theta=\pi/2$ spectrum.

\subsubsection{The resulting one-dimensional energy spectra}
We now consider isotropic, perpendicular and parallel reduced spectra in the light of the spectra studied above. The one-dimensional isotropic total and potential energy spectra, and the parallel and perpendicular spectra, are shown in Fig.~\ref{Ek}. The spectra are averaged in time from the peak of enstrophy until the final time. It was remarked in \cite{cambon_94} that two-dimensional spectra  (Figs.~\ref{cont} and \ref{2Dspec}) may represent a more realistic diagnostics of anisotropic flows given the wide variety of spectral slopes they display as a function of the angle and as a function of the imposed stratification. While this is clear from the previous analysis, the Ozmidov scale and the buoyancy scale will show up more clearly in some of the reduced spectra [in particular, for the buoyancy
 scale, in the potential energy spectrum $E_P(k_{\parallel})$].

Figure \ref{Ek}(a) shows the isotropic total energy spectrum for both runs. The spectrum displays a peak associated with the forcing wavenumber, followed by a flat range (specially in the run with $N=12$).
This flat range extends until a wavenumber close to $k_B$ (see Table \ref{scales}). Then, the run with $N=12$ shows a steep spectrum compatible with $\sim k^{-3}$, while the run with $N=4$ has a short steep range followed by an incipient range compatible with $\sim k^{-5/3}$ after $k_\textrm{oz}^*$. The flat spectrum at large scales is also visible in the 
 parallel spectrum $E_T(k_{\parallel})$ in Fig.~\ref{Ek}(b), but is more evident in the parallel spectrum of potential energy in Fig.~\ref{Ek}(d). On the other hand, the perpendicular spectra in Fig.~\ref{Ek}(c) and Fig.~\ref{Ek}(d) are consistent with $\sim k^{-5/3}$ at all scales and independently of the stratification, as observed before in \cite{lindborg_07}.

The flatness of the spectra at large scales is due to the combination of two related factors: (i) the dominance of the $k_{\perp}=0$ modes observed in Fig.~\ref{2Dspec}, and (ii) the organization of the flow in the vertical direction in well-defined strata with strong vertical gradients both in the velocity and in the buoyancy. It was shown, for example in \cite{kimura_12}, that a superposition of such strata can indeed lead to a flat spectrum since, at large scale, these layers can be interpreted as quasi-discontinuities.

Moreover, the scale at which
 this flat spectrum ends seems to depend linearly with the Froude number, at least for our two runs. The buoyancy scale is generally understood in the context of theoretical studies (see, e.g., \cite{billant_01}) by advocating that the development of turbulence in the vertical direction leads to an effective vertical Froude number $Fr_z=u_\textrm{rms}/[L_B^* N]$ of order unity. In a different context, the buoyancy wavenumber was introduced before in \cite{weinstock_78} to take into account the fact that, in the Lagrangian framework, the buoyancy field is advected by the velocity (although not as a passive scalar) and thus should depend on the total kinetic energy. This leads to the prediction of a sharp break in the buoyancy flux spectrum $\left<w\theta\right>$ at $k_B$, a break that should not develop in the kinetic energy spectrum.

As one moves in the spectra in Fig.~\ref{Ek} to larger wavenumbers, the layers begin to be resolved and their intrinsic dynamics arises. There, the so-called saturation develops; it corresponds to a balance in the vertical between nonlinear advection and buoyancy and leads to 
the $\sim k^{-3}$ spectrum which is more clear for the strongly stratified run ($N=12$). For the weaker stratification the saturation spectrum does not have enough scales to develop and instead one observes a spectrum steeper than the Kolmogorov spectrum in a short range of wavenumbers. Then the collapse of the anisotropic spectra for different $\Theta$ explains the shallower and Kolmogorov-like spectrum for wavenumbers larger than $k_\textrm{oz}^*$ in $E(k)$ for the run with $N=4$. Note that using large-eddy simulations, the transition from a steep (saturated) large-scale spectrum to a Kolmogorov isotropic spectrum was observed before in \cite{carnevale_01} but only sporadically, when breaking events occurred and the turbulence was thus more vigorous.

\begin{figure} \centering
\resizebox{8.5cm}{!}{\includegraphics{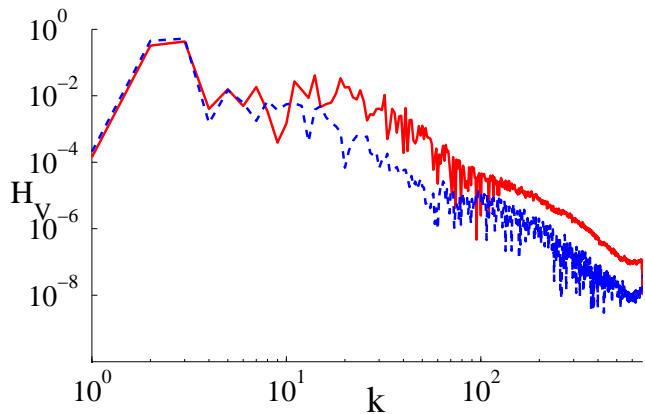}}
\caption{\label{Hk}
{\it (Color  online)} Spectrum of the absolute value of helicity. Solid (red) lines correspond to the run with $Fr\approx 0.03$ ($N=12$ run), while dashed 
(blue) lines correspond to the run with $Fr\approx 0.1$ ($N=4$). Note the flat spectra at large scale, up to what can be identified as the buoyancy scale.} \end{figure}

Finally, Fig.~\ref{Hk} shows the spectrum of the absolute value of helicity for both simulations. Although helicity here is rather small and not important for the flow dynamics, the spectra display, as in the decay runs in Ref.~\cite{rorai_13}, a flat region at large scale followed by a decay at smaller scales with a break near the buoyancy scale. There are rapid changes of sign in the small scales, manifesting as large fluctuations. Interestingly, flat helicity spectra in the planetary boundary layer have been observed at night when the flow is more stably stratified \cite{koprov_05}.

\subsection{Energy fluxes\label{sec:fluxes}}
\begin{figure}[h!tbp] \centering
\resizebox{8.5cm}{!}{\includegraphics{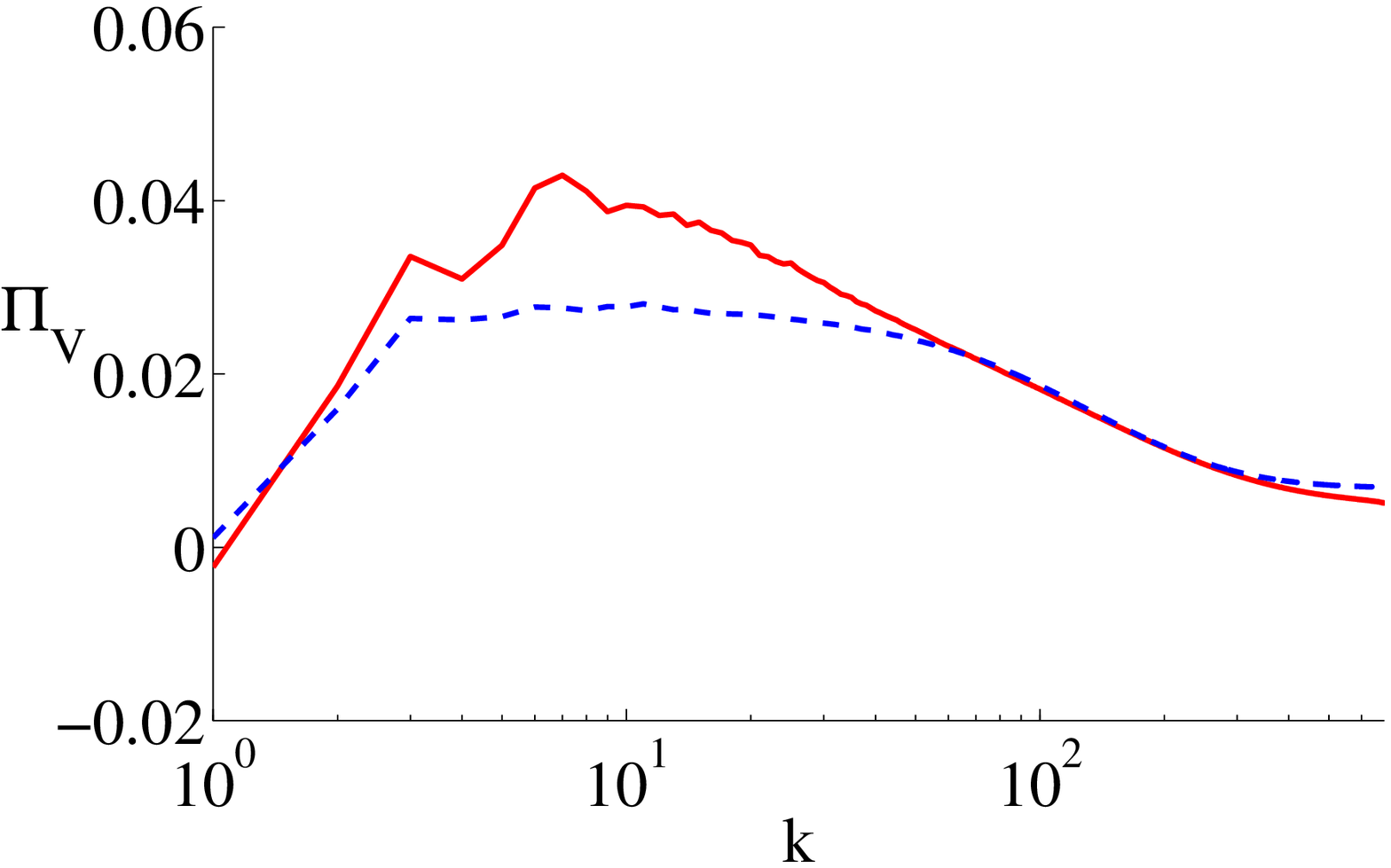}}
\resizebox{8.5cm}{!}{\includegraphics{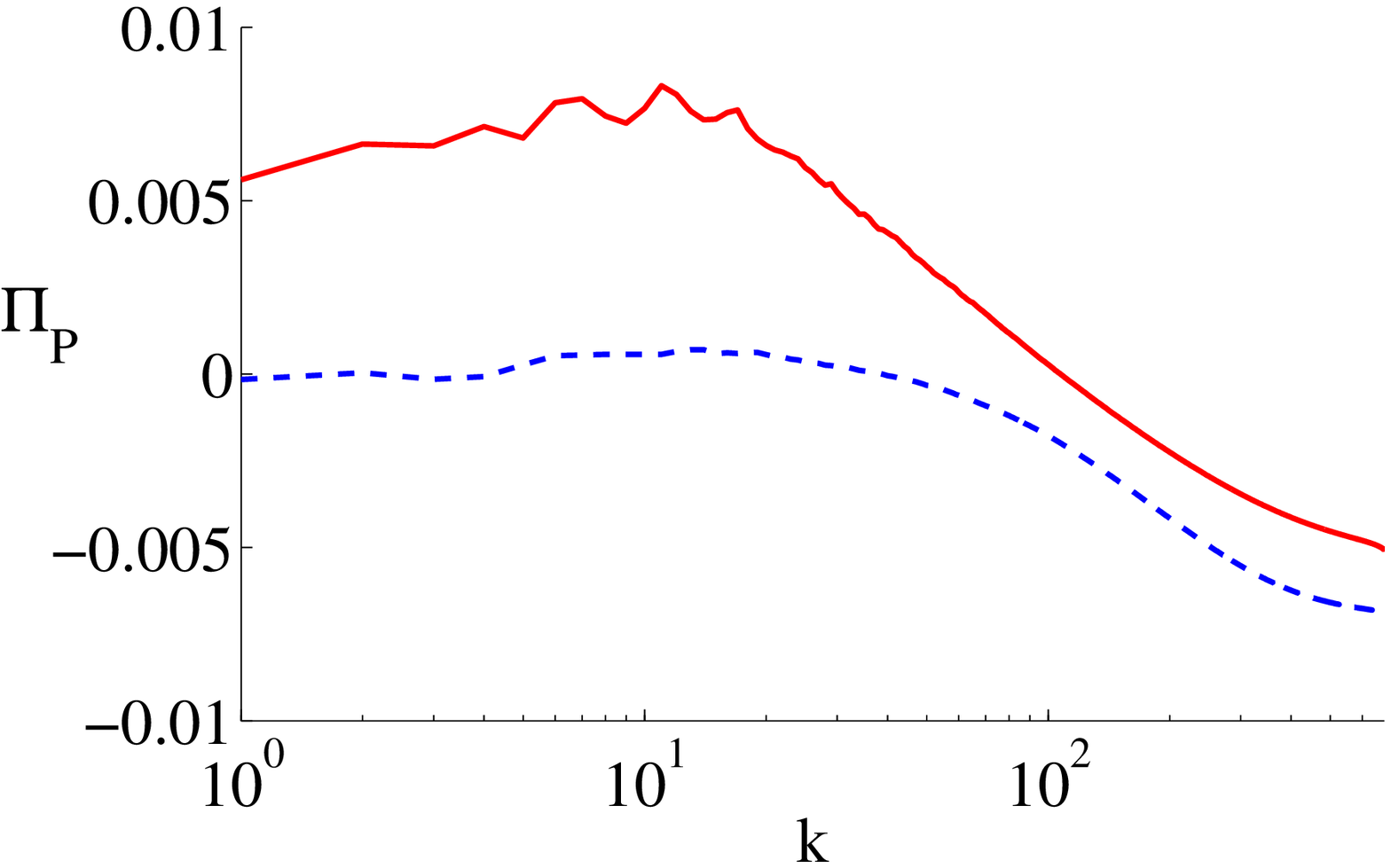}}
\resizebox{8.5cm}{!}{\includegraphics{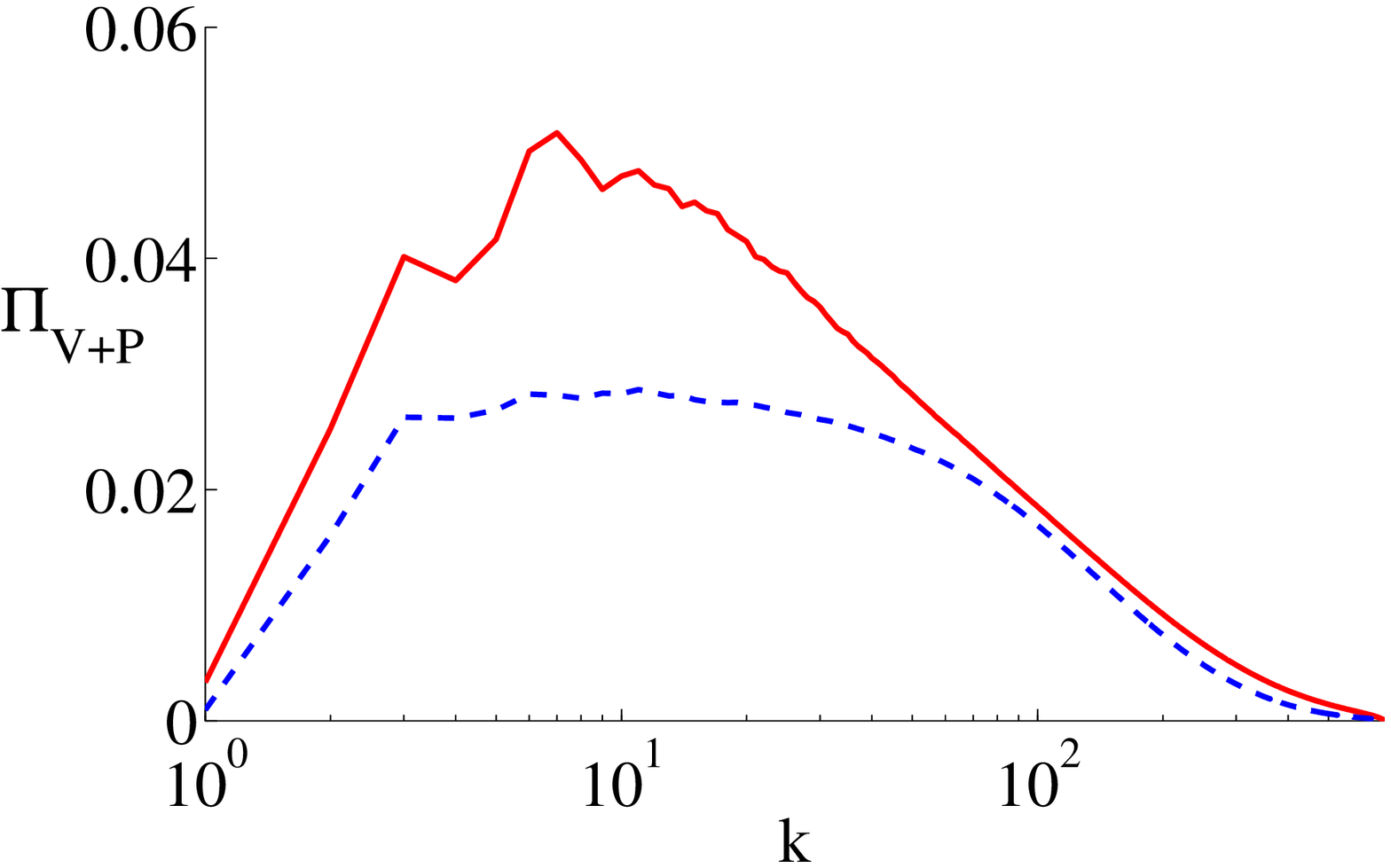}}
\caption{\label{PIVPT}
{\it (Color  online)}  Fluxes of kinetic (top), potential (middle) and total (bottom) energy for the runs with $Fr\approx 0.1$ ($N=4$, indicated by the dashed blue line), and with $Fr\approx 0.03$ ($N=12$, solid red line).} \end{figure}

An examination of energy fluxes confirms the analysis presented in the preceding section. The two simulations, at buoyancy Reynolds numbers of $\approx 27$ and $\approx 220$ respectively, behave differently as to how the energy is being transferred among scales, as can be seen in Fig.~\ref{PIVPT} which displays the kinetic, potential, and total energy fluxes for both runs. Note that the flux constructed from taking only the dot product of Eq.~(\ref{eq:mom}) with the velocity (i.e., the ``kinetic energy flux'') is not a flux, in the sense that its divergence is not zero (i.e., kinetic energy is not conserved alone). Instead, this ``flux'' should be interpreted as energy flux plus power: when it is larger than zero, kinetic energy is transferred towards smaller scales by the velocity field, or injected per unit of time by work done by the temperature. The same applies to the ``potential energy flux'' constructed from dotting Eq.~(\ref{eq:temp}) with the temperature fluctuations: when it is positive, potential energy is transferred towards smaller scales or injected by work done by the velocity, while when it is negative potential energy may be removed by work done by the velocity. Only the total energy defines a proper flux, in the sense that its sign is solely associated with direction of transfer across scales, and in the sense that it goes to zero for $k\to \infty$ (i.e., the total energy is conserved).

In the less stratified run with $N=4$ (high buoyancy Reynolds number ${\cal R}_B$), the total energy flux is approximately constant in a range of wavenumbers that in fact defines the inertial range, with amplitude $\approx 2.7 \times 10^{-2}$. The potential and kinetic energy flux, in the light of the total flux, then indicate how energy is exchanged between the velocity field and the temperature. The potential energy flux is zero at large scale and rather small ($\approx 5 \times 10^{-4}$, or roughly 2\% of the kinetic energy flux) in the same inertial range. It becomes negative and progressively larger (in absolute value) at small scale (after $k\approx 40$), at the end of the inertial range and for $k=k_\textrm{max}$, it reaches $\approx -7 \times 10^{-3}$, a value compensating the kinetic flux at that wavenumber, a condition necessary for energy conservation. The negative value of this flux at small scales indicates that energy is transferred from the small scale temperature fluctuations to the velocity field fluctuations (or in other words, that the small scale temperature gradients exert work on the velocity field, exciting small scale motions). This is in good agreement with the evolution of the enstrophies observed in Fig.~\ref{eta_t}: more energy is dissipated by small scale velocity fluctuations (i.e., by the kinetic enstrophy) than by temperature fluctuations (whose dissipation is associated with the potential enstrophy). Energy at small scales then is transferred from the temperature to the velocity, where it is finally dissipated.

The dynamics of energetic exchanges is rather different at low ${\cal R}_B$. Although the same trends are observed, there is barely a range where the total energy flux is constant; furthermore, all three fluxes are larger in amplitude, but the ratio of kinetic to potential flux is now only roughly equal to 5 at large scales and the potential flux becomes negative at a higher wavenumber $(\approx 100$). One is led to the conclusion that, at that Froude number (and buoyancy Reynolds number), the flow is not sufficiently turbulent even though it produces strong gradients in the vertical.

\section{Discussion and conclusion} \label{s:conclu}
We performed high resolution direct numerical simulations of stratified turbulence for  Reynolds number equal to $Re\approx25000$ and two different Froude numbers: $Fr\approx0.1$ and $Fr\approx0.03$, corresponding to different stratification strengths. Stratified turbulence is modeled through the Boussinesq equations integrated numerically in a three-periodic cubical domain of volume $V = (2\pi)^3$, and discretized with an isotropic grid of $2048^3$ points. The flow is forced at large scale ($k_F = 2$ and $3$) by a three-dimensional randomly generated forcing. By contrasting the behavior of the two simulations we identify some similarities despite the fact that the buoyancy Reynolds number differs by almost an order of magnitude: after an initial transient  the two runs have comparable values of the kinetic and potential enstrophy ($Z_V$, $Z_P$), and energy injection rates ($\varepsilon_V$).  The same ratio  between potential and total energy ($E_P/E_T\approx 0.1$) is spontaneously selected by the flows. For both values of $Fr$, slow modes grow monotonically  as a consequence of nonlinear interactions and cause a slow increase of the total energy ($E_T$) in time. The axisymmetric kinetic energy spectrum, $e_V(k_\perp, k_\parallel)$, clearly shows the anisotropy of the flow, which survives at small scales for the $Fr\approx0.03$ run but not for the simulation at weaker stratification. The axisymmetric total energy spectrum, $e_V+e_\theta$, shows a wide variety of spectral slopes as a function of the angle between the imposed stratification and the wavevector, and with a clear dominance of the slow modes. As a result, the isotropic total energy spectrum is ambiguous because of the superposition of these different dynamical regimes. One-dimensional energy spectra computed in the direction parallel to gravity are flat from the forcing until the buoyancy scale $k_B$. At intermediate scales, a $k_\parallel^{-3}$ parallel spectrum, consistent with the simple 1D model presented in \cite{rorai_14}, develops for the $Fr\approx 0.03$ run, whereas for $Fr\approx 0.1$ the saturation spectrum does not have enough scales to develop and instead one observes a larger slope compatible with a Kolmogorov spectrum $k_\parallel^{-5/3}$. Finally, the spectrum of helicity (velocity-vorticity correlations) is rather weak, but behaves as observed in decaying simulations in \cite{rorai_13}, with its distribution among scales being flat until $L_B$, as also observed in the nocturnal planetary boundary layer.

As observed before in the literature, the dynamics of stratified turbulence proves to be more complex than the homogeneous isotropic case, specially at values of the buoyancy Reynolds number that are intermediate, when waves and eddies strongly interact. Further studies are needed, in particular because there are several relevant scales that must be separately resolved. One issue concerns the effect that the choice of forcing can have on the outcome of the simulations which is far from evident (see for example the discussion in \cite{carnevale_01}). The anisotropic development of large-scales and the ensuing lack of inverse cascade was analyzed in \cite{marino_14}, whereas in this paper we deal with the small-scale anisotropy. However, in these studies
 isotropic forcing (and initial conditions) were used not to bias the development of angular variations. Differences may arise when other forcings, or when correlations between the temperature and the velocity field, are 
 imposed.

Another set of issues is related to the difficulty to perform experiments, in the laboratory as well as numerically, for a set of parameters that accommodates the vast range of physical conditions found in geophysical and astrophysical flows. Part of the difficulty in reaching a full understanding of the behavior of stratified turbulence is the fact that there are different regimes in competition, and that for realistic parameter values for flows in geophysical and astrophysical fluid dynamics, the buoyancy Reynolds number must be sufficiently high, a feature difficult to realize numerically at small Froude number \cite{bartello_13}. But how much more resolution is needed? Spanning the whole range of multi-scale interactions from the largest scale to the dissipative scale, to cover a potential inverse transfer feeding the slow modes \cite{waleffe_93, smith_99, smith_02, marino_14}, a range where energy is fed into the system, a range dominated by wave interactions, followed by a range dominated  by nonlinear eddies, and finally a dissipative range is currently impossible, and various choices have been made in the past even when computing at high resolutions up to $8192^3$ grid points \cite{debruynkops_14} (see also \cite{brethouwer_09,  3072, bartello_13}). Resort to high-performance computing at higher resolution, as well as to modeling, will be some of the avenues to be followed in the near future.

{\it This work was supported by NSF/CMG 1025183. C. Rorai acknowledges support from two RSVP/CISL grants. Computer time on Yellowstone through an ASD allocation was provided by NCAR under sponsorship of NSF.}

\bibliography{ms_14_v2.bbl}
\end{document}